\documentstyle[preprint,aps]{revtex}                                   
\draft        
\begin{document}             
\title{Zeroth Sound Modes of Dilute Fermi Gas with Arbitrary Spin}
\author{S-K. Yip \cite{Taiwan} and Tin-Lun Ho}
\address{Physics Department, The Ohio State University,
Columbus, OH 43210}
\maketitle          

\maketitle

\begin{abstract}
Motivated by the recent success of optical trapping of alkali bosons, we have 
studied the zeroth sound modes of dilute Fermi gases with 
arbitrary spin-$f$, which are spin-$S$ excitations ($0\leq S\leq 2f$). 
The dispersion of the mode $(S)$ depends on a single Landau  parameter 
$F^{(S)}$, which is related to the scattering lengths of the system 
through a simple formula.  Measurement of (even a subset of)  
these modes in finite magnetic fields will enable one to determine
all the interaction parameters of the system.
\end{abstract}

Since the discovery of Bose-Einstein condensation (BEC) in dilute gases of 
alkali atoms\cite{BEC}, there have been experimental efforts to cool alkali 
fermions such $^{6}$Li and $^{40}$K down to the degenerate limit. In current 
experiments, alkali atoms are confined in magnetic traps, which 
confine only the 
spin states aligned with the local magnetic field. As a result, 
the spinor nature of the atom is suppressed. The recent success of    
optical trapping\cite{MIT98}, however, changes the situation.
In optical traps,  different spin components are degenerate 
in the absence of magnetic fields. One therefore has the opportunity 
to study dilute Bose gases with  integer hyperfine spins (or 
simply ``spin") $f>0$ and Fermi gases with spins $f>1/2$.
In a recent paper\cite{HoYip}, we have discussed the structure of 
Cooper pairs of alkali fermions in
optical traps. Since most alkali fermions have spin $f>1/2$, 
their Cooper pairs can have {\em even} spin $J$ ranging from 0 to $2f-1$. 
The internal 
structures of these large spin Cooper pairs will give rise of 
to a great variety of superfluid phenomena. 

The purpose of this paper is to study a key {\em normal} state property 
of dilute fermi gases with general spin-$f$ in the degenerate limit, $--$
their collisionless or ``zeroth" sound.
We shall show that in addition to the ordinary density mode, the system has 
additional modes corresponding 
to coherent inter-conversions of different spin species. These modes are the 
generalizations of the spin waves of spin $1/2$ Fermi liquids. 
As we shall see, the dispersions of the zeroth sound modes contain the
information of {\em all} the interaction parameters of the system, i.e.
the set of s-wave scattering lengths $\{ a_{_J}\}$ of two spin-$f$ atoms in the 
total spin $J$ channel. Thus,    
observation of these modes  
will not only provide evidence of the degenerate nature of 
the system, but also information about the scattering lengths 
$a_{_J}$, and hence the existence of superfluid ground state 
as well as their transition temperatures. 

As in our previous study\cite{HoYip}, 
we shall focus on the homogenous case, i.e. without external potential. 
This is
a necessary step before studying trapped fermions. Moreover, it is 
conceivable 
that optical traps of the form of cylindrical {\em boxes} (rather 
than harmonic wells) be constructed in the future. In that case, the  
discussions here will be directly applicable. As in our previous  
work\cite{HoYip},  our symmetry classification of the spin structure 
(which is a crucial step in our solution) also applies to arbitrary
potentials. 

In addition to homogeneity, we shall also consider the 
weak magnetic field limit, i.e. when the Zeeman energy is much smaller than the 
kinetic energy of the system. 
These are the regimes where the 
spinor nature of the fermi gas is manifested most clearly. 
As demonstrated by the recent experiments at MIT\cite{Stenger},
this limit can be easily achieved by specifying the 
total spin of the system. Since  
the low energy dynamics of the system is spin conserving\cite{Hospin1},
the specified spin can not relax.  The system therefore sees an effective
magnetic field with which its spin would be in equilibrium, a field which 
can be  much smaller than the external field $B^{\rm ext}$. 
In the following, we shall 
refer to this effective field simply as ``magnetic field" $B$, with the 
understanding that it is a Lagrange multiplier that determines the total 
spin of the system\cite{Stenger}.

{\bf (A)} {\em Zero magnetic field :}
We begin with the  linearized kinetic equation for the 
distribution function matrix $\delta \hat{n}_{\bf p}$ in the 
collisionless regime
\begin{equation}
{\partial \delta \hat{n}_{\bf p}({\bf r},t) \over \partial t} 
+ {\bf v}_{\bf p}\cdot{\bf \nabla}\left( 
 \delta \hat{n}_{\bf p}({\bf r},t) - \frac{\partial n^{o}_{\bf p}}{\partial 
\epsilon_{\bf p}}  
   \delta \hat{\epsilon}_{\bf p}({\bf r}, t) \right)   
     = 0 , \label{ke1}  
\end{equation}
\noindent  Our notations in eq.(\ref{ke1})  are the same as ref.\cite{LL}. 
Here,  $n^{o}_{\bf p}$ is the $T=0$ Fermi function, 
${\bf v}_{\bf p}= \nabla_{\bf p}\epsilon_{\bf p}$, $\delta \hat{n}_{\bf p}$  
is a $(2 f + 1)  \times (2 f + 1) $ matrix 
in spin-space,  $[\delta \hat{n}_{\bf p}({\bf r},t)]_{\alpha\beta}$$=
\int {\rm d}{\bf x}e^{-i{\bf p}\cdot{\bf x}}<\psi^{+}_{\beta}
({\bf r}-{\bf x}/2,t)
\psi_{\alpha}({\bf r}+{\bf x}/2,t)>$
where $\psi_{\alpha}$ is the field operator. 
The energy matrix $\delta \hat \epsilon$ describes the change in the 
Hamiltonian due to $\delta \hat{n}$, 
\begin{equation}
 [\delta \epsilon_{\bf p}]_{\alpha\beta}= \int {\rm d}\tau' 
 f_{\alpha\gamma, \beta\delta} ({\bf p}, {\bf p}')
[\delta n_{{\bf p}'}]_{\delta\gamma} 
 \label{deff}
\end{equation}
where ${\rm d}\tau'$ means ${\rm d}{\bf p}'/(2\pi)^{3}$, and 
 $f_{\alpha\gamma, \beta\delta}({\bf p}, {\bf p}')$ are the 
Landau parameters which can be extracted from the  
Hamiltonian of the system derived by one of us\cite{Hospin1}.  
It is shown in ref.\cite{Hospin1} that only the lowest hyperfine
states (with spin $f$) will remain in the optical trap and that 
the interactions 
between these spin-$f$ atoms are spin conserving, of the  form 
\begin{equation} 
H_{int} = \frac{1}{2}\sum_{J=0,2,..}^{2f-1} g_{_J}\sum_{M=-J}^{J}
\int {\rm d} {\bf r} O^{+}_{JM}({\bf r})O_{JM}({\bf r}), 
\label{ham} \end{equation}
where $O_{JM}({\bf r}) = \sum_{\alpha\beta}< JM \vert ff\alpha\beta>
\psi_{\alpha}({\bf r})\psi_{\beta}({\bf r})$, and 
$<JM \vert f_{1}f_{2} \alpha\beta>$ are  the Clebsch-Gordan
coefficients for forming a spin-$J$ object from a spin-$f_{1}$ and a
spin-$f_{2}$ particle\cite{short}, 
$g_{_J}=4\pi\hbar^{2} a_{_J}/M_F$, and $M_F$ is the mass 
of the atom. 
Pauli principle implies that only even $J$'s exist in eq.(\ref{ham}). 
Evaluating $<H_{int}>$ in Hartree-Fock approximation, and using 
the fact that 
$\delta<\psi^{+}_{\alpha}({\bf r})\psi_{\beta}({\bf r})>
= \int {\rm d}\tau [\delta n_{\bf p}({\bf r})]_{\beta\alpha}$, it is 
straightforward to show that 
\begin{equation}
   f_{\alpha\gamma, \beta\delta} =  2\sum_{J=0,2, ..}^{2f-1} g_{_J}
\sum_{M=-J}^{J}
 < JM\vert ff\gamma\alpha>< JM\vert ff \delta \beta>, 
     \label{f}
\end{equation}
which is momentum independent as a result of s-wave interaction. 
Note that  if $g_{_J}< 0$, the system will have a superfluid instability
towards  spin-$J$ Cooper pairs at a sufficiently low
temperature $T_{c}^{(J)}$\cite{HoYip}. Our discussions for negative $g_{_J}$'s
therefore applies to temperatures above  $T_{c}^{(J)}$  but  low 
enough  so that 
the Fermi gas is degenerate. Before proceeding, we simplify 
eq.(\ref{ke1})  by writing
$\delta \hat{n}_{\bf p}=  \left( -\frac{\partial n^{o}_{\bf p}}
{\partial \epsilon_{\bf p}}\right)
\hat{\nu}_{\hat{\bf p}}$, which turns eq.(\ref{ke1}) and (\ref{deff}) into 
\begin{equation}
\partial_{t}\hat{\nu}_{\hat{\bf p}} + 
{\bf v}_{\bf p}\cdot {\bf \nabla} \left( \hat{\nu}_{\hat{\bf p}} + \delta 
\hat{\epsilon}_{\hat{\bf p}} \right) =0 , \,\,\,\,\,\,\,\,\,
\delta \hat{\epsilon}_{\hat{\bf p}} = N_{F} f_{\alpha\gamma, \beta\delta} 
<\hat{\nu}_{\hat{\bf p}}> , 
\label{kenu} \end{equation} 
where $N_F = mk_{F}/2\pi^{2}\hbar^{2}$ is the density of state of a single 
spin component at the Fermi surface, 
$k_{F}$ is the Fermi wavevector, and $<(..)>\equiv \int 
\frac{ {\rm d}\hat{\bf p} }{4\pi} (..) $ denotes the angular 
average over the Fermi 
surface.  Note that the quasi particle energy  
$\delta \hat{\epsilon}_{\hat{\bf p}}$ is
isotropic in k-space as a consequence of the s-wave interactions between 
the particles. 

Next, we note that a rotation $\vec{\theta}$ in spin space will cause a 
change $a_{\alpha}\rightarrow D^{(f)}_{\alpha\beta}(\vec{\theta})a_{\beta}$,
where $D^{(f)}_{\alpha\beta}$ is the rotation matrix in the spin-$f$ space. 
This implies $\hat{\nu} \rightarrow  \hat{\nu}' = 
\hat{D}^{(f)}\hat{\nu} \hat{D}^{(f)+}$. 
>From eq.(\ref{deff}) and (\ref{f}),  one can see that 
$\delta \hat{\epsilon}_{\bf p}$ transforms the same way, 
$\delta \hat{\epsilon}_{\bf p}
\rightarrow \hat{D}^{(f)} \delta \hat{\epsilon}_{\bf p}  \hat{D}^{(f)+}$.
Since $\hat{\nu}$ is made up of two spin-$f$ objects, it can be 
decomposed into a sum of spin-$S$ quantities $\hat{\nu}^{(S,M)}$,
which transform as $[\hat{D}^{(f)}\hat{\nu}^{(S, M)} 
\hat{D}^{(f)+}]_{\alpha\beta} = 
[ \hat{\nu}^{(S,M')}]_{\alpha\beta} D^{(S)}_{M'M}$, 
where $0\leq S \leq 2f$, $-S\leq M \leq S$. 
The solution of this equation is easily seen to be 
$[\delta \hat{n}^{(S,M)}]_{\alpha\beta} \propto <f\alpha\vert Sf M\beta>$. 
We then have the representation 
\begin{equation}
\left( \begin{array}{c} \left[\nu_{\bf p}({\bf r}, t) \right]_{\alpha\beta} \\
 \left[ \delta \epsilon_{\bf p}({\bf r}, t)\right] _{\alpha\beta} 
\end{array} \right)
= \sum_{S,M}  <f\alpha\vert Sf M\beta>  
\left( \begin{array}{c} \nu^{(S,M)}_{\hat{\bf p}} ({\bf r}, t) \\
\delta \epsilon^{(S,M)}_{\hat{\bf p}}({\bf r}, t) \end{array} \right)
\label{rep} \end{equation}
Substituting eq.(\ref{rep}) into eq.(\ref{kenu}) and using the
identity
\begin{eqnarray}
\sum_{\gamma\delta M'}<JM'\vert ff\gamma\alpha><JM'\vert ff\delta \beta>
<f\delta\vert SfM\gamma>  \nonumber \\
 =   (-)^{2f-J}(2J+1)W(ffff; JS)<f\alpha\vert SfM\beta>
\end{eqnarray}
where $W$ is the Racah coefficient\cite{Brink}, eq.(\ref{kenu}) becomes 
diagonal in the $(S,M)$ modes, 
\begin{equation}
\partial_{t}\hat{\nu}^{(S,M)}_{\hat{\bf p}} 
+ {\bf v}_{\bf p}\cdot{\bf \nabla}\left( \hat{\nu}^{(S,M)}_{\hat{\bf p}} 
+ \delta \hat{\epsilon}^{(S,M)}_{\hat{\bf p}} \right) =0, 
\label{nuSM} \end{equation}
\begin{equation}
\delta \epsilon^{(S,M)}_{\hat{\bf p}} = F^{(S)}<\nu^{(S,M)}_{\hat{\bf p}}>, 
\label{enSM}  \end{equation}
\begin{equation}
F^{(S)}= - \sum_{J=0,2,...}^{2f-1} \frac{4k_{F} a_{_J}}{\pi} 
(2J+1) W(ffff; JS) , 
\label{FS} \end{equation}
where we have used the fact 
that $N_{F}g_{_J}= 2k_{F}a_{_J}/\pi$ 
and $(-1)^{2f-J} = -1$ in obtaining eq.(\ref{FS}).
Eq.(\ref{nuSM}) and (\ref{enSM}) imply that 
\begin{equation}
\left( \frac{\partial}{\partial t} + {\bf v}_{\bf p}\cdot{\nabla} \right)
\nu^{(S,M)}_{\hat{\bf p}} +  F^{(S)} {\bf v}_{\bf p}\cdot{\nabla}
<\nu^{(S,M)}_{\hat{\bf p}}> =0 . 
\label{ke2} \end{equation}
which is precisely the equation for the ordinary zeroth sound mode with only 
$\ell=0$ spin-symmetric  
Landau parameter  $F^{s}_{\ell=0}$ non-zero\cite{LL} and is given by 
$F^{(S)}$. 

The dispersion relations of modes described by eq.(\ref{ke2}) is 
well known\cite{LL}. They are 
\begin{equation}
1 = F^{(S)} \int^{1}_{-1} \frac{{\rm d}x}{2} \frac{ qv_{F} x}{\omega - qv_{F}x } .
\label{dis} \end{equation}
The properties of the modes
depend crucially on the sign of the parameters $F^{(S)}$. 
When $F^{(S)}>0$, one has a well defined propagating mode. 
When $-1<F^{(S)}<0$, the zeroth sound mode is Landau damped. When 
$F^{(S)}<-1$, the system is unstable against spin-$S$ distortions. 
Because of the dilute limit, $k_{F}a<<1$ and hence $|F^{(S)}|<1$, 
stability against spin-$S$ distortions
is guaranteed.      

It is instructive to consider some special cases : 

\noindent {\bf (i)} {\em The density modes $\hat{\nu}^{(S=0)}$ 
for fermions with arbitrary 
spin-$f$} : Using the fact that 
$W(ffff; J0)= (-1)^{2f-J}/(2f +1)$, we have 
\begin{equation}
F^{(S=0)} = \frac{4}{\pi(2f+1)} \sum_{J=0,2, ..}^{2f-1} (2J+1) k_{F}a_{_J} 
 \label{F0} \end{equation}
In particular, if there are no superfluid instabilities in all angular momentum 
$J$ channel,  then $F^{(S=0)} > 0$ and 
the density mode will not be Landau damped. 

\noindent {\bf (ii)} {\em Fermions with spin-$1/2, 3/2$ and $5/2$} :   
 For $f= 1/2$, eq.(\ref{F0}) reduces to the well-known results
 $F^{(S=0)} = -F^{(S=1)} = N_F g_0 = 2 k_F a_0 / \pi \hbar$\cite{LL}.
 For $f=3/2$,  using the tabulated values of the $6j$-symbols\cite{Rotenberg} 
to calculate the Racah coefficients, we find 
 $F^{(S=0)} = k_F (a_0 + 5 a_2)/\pi$,
 $F^{(S=1)} = F^{(S=3)} = - k_F (a_0 + a_2)/\pi$, 
 and $F^{(S=2)} = k_F (a_0 -3 a_2)/\pi$.  Thus the $S=1$ and $3$ modes are
 always degenerate, and the degeneracy between the $S=0$ and $S=2$ 
 modes are lifted only by the interaction in the $J=2$ 
 channel.  If there are no superfluid instabilities in any $J$ channel, 
i.e. both $a_{0}, a_{2}>0$,  then $S=1$ and $S=3$ modes
 are always Landau damped.

 For large $f$'s there are no obvious systematics except
 for the $S=0$ result noted above.  
 Modes for different $S$'s are typically not degenerate,
barring accidental values of $g_{_J}$'s.  
In the case of $f=5/2$ such as $^{22}$Na and $^{86}$Rb, we obtain
$F^{(0)} =  ( 2a_0/3 + 10 a_2/3 + 6 a_4 )k_F/\pi$, 
$F^{(1)} = -(  2a_0/3 + 46 a_2/21 - 6 a_4/7 )k_F/\pi$, 
$F^{(2)} = ( 2a_0/3 +  a_2/3 -3 a_4 )k_F/\pi$, 
$F^{(3)} = - ( 2a_0/3 -29 a_2/21 + 19 a_4/7 )k_F/\pi$, 
$F^{(4)} = ( 2a_0/3 -5 a_2/3 - a_4 )k_F/\pi$, 
$F^{(5)} = - ( 2a_0/3 + 25 a_2/21 +  a_4/7 )k_F/\pi$. 

\noindent {\bf (iii)} {\em Undamped zeroth sound modes in the dilute limit} : 
Zeroth sound modes $(S,M)$ with $F^{(S)}>0$ are propagating\cite{LL}. Because 
of the  dilute condition  $k_{F}a_{_J}<<1$, we have $F^{(S)}<<1$. 
In this limit, eq.(\ref{dis})
can be integrated to give\cite{LL}, 
\begin{equation}
\omega^{(S)}(q) = qv_F \left(1 + 2e^{-2} e^{-2/F^{(S)}}\right)  .
\label{omegas}
\end{equation}
Since $F^{(S)}<<1$, the exponential term in eq.(\ref{omegas}) will have little 
contributions. The frequencies of zeroth sound for all $S$ are essentially 
given by $qv_F$. As a result, it will be difficult to obtain  
information of the interaction parameters from zeroth sound frequencies 
in zero field. 
On the other hand, we shall see that even a small magnetic field will 
cause significant changes in the zeroth sound dispersions, 
which lead to many observable features and enable one to determine all 
the  interaction parameters.

{\bf (B)} {\em Weak magnetic fields:} When $B\neq 0$, the kinetic equation 
eq.(\ref{ke1}) will have an additional term ${\cal I}_{\bf p}= \frac{i}{\hbar}
[\hat{\epsilon}_{\bf p}, \hat{n}_{\bf p}]$ on the left hand 
side\cite{LL}. At the same time,
 the {\em equilibrium} 
distribution function and quasiparticle energy (denoted as 
$\hat{n}^{o}_{{\bf p},B}$ and 
$\hat{\epsilon}^{o}_{{\bf p},B}$ respectively) are altered from the zero 
field values ($\hat{n}^{o}_{\bf p}$ and  $\hat{\epsilon}^{o}_{\bf p}$).   
The difference $\delta \hat{\epsilon}^{o}_{\bf p} = 
\hat{\epsilon}^{o}_{{\bf p},B} - \hat{\epsilon}^{o}_{\bf p}$ is 
$[\delta \epsilon^{o}_{\bf p}]_{\alpha\beta}$$= 
-\mu B F^{z}_{\alpha\beta}+ 
\int {\rm d}\tau'f_{\alpha\gamma, \beta\delta}[\delta
n^{o}_{\bf p'}]_{\delta\gamma}$, with 
$\delta \hat{n}^{o}_{\bf p}$$=\hat{n}^{o}_{{\bf p}, B}$
$-\hat{n}^{o}_{\bf p}$$=\left(\frac{\partial n^{o}}
{\partial \epsilon_{\bf p}}\right) \delta \hat{\epsilon}^{o}_{\bf p}$. 
These two relations imply 
\begin{equation}
[\delta \epsilon^{o}_{\bf p}]_{\alpha\beta} = -\mu B F^{z}_{\alpha\beta} -
N_{F}f_{\alpha\gamma, \beta\delta}
 [\delta \epsilon^{o}_{\bf p}]_{\delta\gamma} , 
\label{equil} \end{equation}
where $\mu$ is the magnetic moment of the atom. 
The solution of eq.(\ref{equil}) is $\delta\hat{\epsilon}^{o}_{\bf p} 
= c\hat{F}^{z}$. 
Using the fact that $(F^{z})_{\alpha\beta} \propto <f\alpha\vert 1f0\beta>$, 
it is easily to show from eq.(\ref{rep}), (\ref{FS}), and 
(\ref{equil}) that 
\begin{equation}
\delta\hat{\epsilon}^{o}_{\bf p} = - \mu B^{\rm eff} \hat{F}^{z}, \,\,\,\,\,\,\,
B^{\rm eff}= B/(1+F^{(1)}) . 
\label{Beff} \end{equation}

Linearizing about the equilibrium configuration 
$\hat{n}^{o}_{{\bf p},B}$ and $\hat{\epsilon}^{o}_{{\bf p},B}$, we have 
${\cal I}_{\bf p} = \frac{i}{\hbar}\left([\delta \hat{\epsilon}^{o}_{\bf p},
\delta\hat{n}_{\bf p}] + [\delta\hat{\epsilon}_{\bf p}, \delta\hat{n}^{o}_{\bf p}]
\right)$.  
>From the definition $\delta \hat{n}_{\bf p} = -\left( 
\frac{\partial n^{o}_{\bf p}}{\partial \epsilon_{\bf p}}\right) 
\hat{\nu}_{\hat{\bf p}}$, 
the relation $\delta\hat{n}^{o}_{\bf p} = 
\left( \frac{\partial n^{o}_{\bf p}}{\partial \epsilon_{\bf p}}\right)\delta 
\hat{\epsilon}^{o}_{\hat{\bf p}}$, and the property 
$(\alpha-\beta)[\nu_{\bf p}]_{\alpha\beta} = 
\sum_{SM}<f\alpha\vert SfM\beta>M\nu^{(S,M)}_{\hat{\bf p}}$ which follows
from eq,(\ref{rep}),  we have 
\begin{equation}
\left[{\cal I}_{\bf p}\right]_{\alpha\beta}
= \frac{i}{\hbar}\left( \frac{\partial n^{o}}{\partial \epsilon_{\bf p}}\right)
\sum_{SM}<f\alpha\vert SfM\beta>(\Omega M)
\left( \nu^{(S,M)}_{\hat{\bf p}} + \delta\epsilon^{(S,M)}_{\hat{\bf p}} \right)
\end{equation}
where $\Omega\equiv \mu B^{\rm eff}/\hbar$. 
With this additional term on the left hand side of eq.(\ref{ke1}) and repeating the 
procedure as before, we have find that eq.(\ref{enSM}) remains unchanged whereas 
eq.(\ref{nuSM}) becomes
\begin{equation}
\partial_{t}\hat{\nu}^{(S,M)}_{\hat{\bf p}}
+ \left[ {\bf v}_{\bf p}\cdot{\bf \nabla} -  i\Omega M\right] 
\left( \hat{\nu}^{(S,M)}_{\hat{\bf p}}
+ \delta \hat{\epsilon}^{(S,M)}_{\hat{\bf p}} \right) =0 . 
\label{nuSMB} \end{equation}
Thus, the zeroth sound modes can still be classified
by the quantum numbers $(S, M)$  in the weak field limit. 
The equation for the dispersion now becomes 
\begin{equation}
1 = F^{(S)} \int^{1}_{-1} \frac{ {\rm d}x}{2} \frac{qv_{F} x  -  \Omega M}
{\omega  +  \Omega M - qv_{F} x} ,  
\end{equation}
which upon integration gives 
\begin{equation}
\frac{1}{F^{(S)}} = \frac{\omega}{2qv_{F}}{\rm ln}
\frac{{\omega} + \Omega M + qv_{F}}
{{\omega} +  \Omega M - qv_{F}} -1,  \,\,\,\,\,\,\,\,
\Omega = \frac{\mu B}{1+F^{(1)}}
\label{dispB2} \end{equation}

Since  the collective modes are excitations above the ground state, we only 
need to study the  $\omega >0$ solutions of eq.(\ref{dispB2}). 
In the following, we shall discuss only the zeroth sound modes that are not
Landau damped, which requires $|\omega + \Omega M|>qv_{F}$ 
in eq.(\ref{dispB2}). 
While many features of these propagating modes can be 
obtained analytically, we first 
display the numerical solutions of eq.(\ref{dispB2}) for $S=3/2$ with 
$F^{(S)}>0$ and $F^{(S)}<0$ in fig.1 and fig.2 respectively. 
The notable features of these modes are : 

\noindent {\bf (iv)} {\em Zeroth Sound modes near $q=0$} :  
Near $q = 0$,  $qv_{F}/|\Omega M| <<F^{(S)}$, it is easily seen from 
eq.(\ref{dispB2}) that\cite{Silin}
\begin{equation}
\omega^{(S,M)}(q) = -\Omega M (1+F^{(S)}) \left[ 1 +
\frac{1}{3F^{(S)}}
\left( \frac{qv_{F}}{\Omega M} \right)^{2}  + ..\right]. 
\label{q2} \end{equation}
Since $\omega^{(S,M)}>0$, only $\Omega M<0$ modes can be excited at $q=0$. 
Note that {\em all 
$q=0$ modes in finite field are not Landau damped 
 irrespective of the sign of $F^{(S)}$}. 
>From eq.(\ref{q2}), one can also see that all $\omega^{(S,-|M|)}$ modes
increase (decrease) as $q^2$ for $F^{(S)}>0$ and $<0$. (See also fig.1 and 2).

\noindent {\bf (v)} {\em The $F^{(S)}>0$ case} : 
For $F^{(S)}>0$, zeroth sound modes with $\Omega M>0$ emerge 
from $\omega =0$ when 
$q>q_{_M}\equiv \Omega M/v_F$. (See fig.1). 
Expanding eq.(\ref{dispB2}) about $(\omega=0,q=q_{M})$, one finds
the dispersion in this neighborhood is 
\begin{equation}
\omega - ( qv_{F} - \Omega M) = 2 \Omega M e^{-\left(1+1/F^{(S)}\right)
\left(2 \Omega M /\omega\right)}
\label{qm} \end{equation}  
Another simple feature one can derive from eq.(\ref{dispB2}) is that 
as $q$ increases so that $qv_F >>|\Omega M|$, 
$\omega^{(S,M)}(q) \to \omega^{(S)}(q) - \Omega M$.
The dispersions for all $(S,M)$ modes become parallel to $\omega = qv_F$, 
with all  $\Omega M<0 (>0)$ modes shifted up (down) by an amount of 
$|M\Omega|$. (See fig.1).
It is also straightforward to show that zeroth sound modes with 
$\Omega M\neq 0$ lie above the particle-hole continuum of that 
particular $M$ state, i.e.  
$\omega > -\Omega M+ qv_{F}$.

\noindent {\bf (vi)}  {\em The $F^{(S)}<0$ case} :
We find that the $\Omega M <0$ modes decrease monotonically   as $q$
increases, and vanish at $q_{_{|M|}}=|\Omega M|/v_F$ in a manner 
similar to eq.(\ref{qm}). The entire mode $M$ lies
below the corresponding particle-hole continuum,
i.e., $\omega < - \Omega M - q v_F$.
Solving eq.(\ref{dispB2}) graphically, 
one can also see that there are 
no solutions with $\omega >0$ when $\Omega M>0$, implying the absence of 
zero sound modes with $\Omega M>0$. 

{\em Determination of scattering lengths} : 
For scattering lengths $|a_{_J}|\sim 100a_{B}$ where $a_{B}$ is the
Bohr radius, a Fermi gas of density $\sim 10^{13}$cm$^{-1}$ will have 
$k_{F} a_{_J}\sim 10^{-1}$, which implies $\{ F^{(S)}\}\sim 10^{-1}$. 
As mentioned in Part ({\bf A}), it will be hard to determine
the scattering lengths $a_{_J}$ from the $B=0$ zero sound modes for these
values of $\{ F^{(S)}\}$ because their small contributions. 
On the other hand, in the presence of magnetic field, different zeroth 
sound modes $(S,M)$ are separated. 
Since the interaction contributions to the zeroth sound frequency at $q=0$ 
and to the critical wavevector $q_{M}$ are of the form $1+F^{(S)}$ instead of 
the essential singularity form in eq.(\ref{omegas}), their contributions should
be measurable for $k_F a_{_J}\sim 10^{-1}$ or even smaller. Note also that
there are only  $(2f+1)/2$ scattering lengths 
$[ a_{_{J}}, (J=0,2, .., 2f-1)]$ whereas the  number of zeroth  sound modes
in finite field is  $\sum_{S=0}^{2f}(2S+1) = (2f+1)^{2}$. Even some of these
modes may not be excited, (as in the case of $F^{(S)}<0$), there are still more
the conditions on $a_{_J}$ provided by the zeroth sound frequencies
than the number of $a_{_J}$ themselves. Thus, it is possible to determine 
the entire set of scattering lengths $\{ a_{_J} \}$ from 
the zero sound dispersions. 

This work is supported by NASA Grant NAG8-1441, 
NSF Grants DMR-9705295 and DMR-9807284.

\vspace{0.2in}

\noindent Figure Captions : 

\vspace{0.2in}

\noindent Figure 1 : The zeroth sound mode for $F^{(2)}=0.5>0$, 
$f$ arbitrary, and $\Omega= +0.2$. From upper to lower, the curves 
correspond to $M=-2,-1,0,1,2$ respectively. 
Both $\omega, qv_F$, and $\Omega$ are plotted with 
arbitrary units.  The vertical intercepts of curves $M=-2$ and $-1$ 
are $\omega^{(2, -2)} = 2\Omega (1+F^{(2)})$ and 
 $\omega^{(2, -1)} = \Omega (1+F^{(2)})$ respectively. 
The horizontal intercepts of the $M=1$ and $M=2$ curves 
are $q_{1}v_{F} = \Omega$ and   $q_{2}v_{F}= 2\Omega$ respectively. 
Zeroth sound modes for other $S$ have different number of branches but behave 
similarly. 

\vspace{0.2in}

\noindent Figure 2:  The zeroth sound mode for the case $F^{(2)}= -0.2<0$, 
$f$ arbitrary. From upper to lower, the curves correspond to $M=-2$ and 
$-1$. The value of $\Omega$ and the  expressions for the vertical 
intercepts are identical to those in figure 1. The horizontal intercepts are 
 $q_{-1}v_{F} = \Omega$ and   $q_{-2}v_{F}= 2\Omega$ respectively.
 

\begin{thebibliography}{10}
\bibitem{Taiwan} Present Address : 
Physics Division, Center for Theoretical Sciences
P. O. Box 2-131, Hsinchu, Taiwan 300, R. O. C.
\bibitem{BEC}  M.H. Anderson, et.al., Science {\bf 269}, 198 (1995).
K. B. Davis, et.al., Phys. Rev. Lett. {\bf 75}, 3969 (1995). 
\bibitem{MIT98} D. M. Stamper-Kurn {\it et al}, {\it Phys. Rev. Lett}
  {\bf 80}, 2027 (1998)
\bibitem{HoYip} T.L. Ho and S.K. Yip, Cond-mat/9808264.
\bibitem{Stenger}  J. Stenger, et.al. {\em Spin domains in ground state spinor
Bose-Einstein condensates}, preprint.
\bibitem{Hospin1}  T.L. Ho, Phys. Rev. Lett. {\bf 81}, 742, (1998).

\bibitem{LL} L. D. Landau and E. M. Lifshitz,
  Statistical Physics II, Pergamon Press, 1981.

\bibitem{short} The proper convention is $<ffFM \vert f f m_1 m_2>$. The 
first $``ff"$ is suppressed since no confusions will arise.
  
\bibitem{Brink} D. M.  Brink and G. R.  Satchler, {\em Angular Momentum},
Oxford University Press, 1968.

\bibitem{Rotenberg} M. Rotenberg et al, {\em The $3-j$ and
$6-j$ symbols}, the Technology Press, MIT 1959,

\bibitem{Silin} The dispersion for the spin wave of spin 1/2 fermions was 
first studied by V.P. Silin, JETP, vol.6, 945, (1958), and later by 
A.J. Leggett, J. of Physics C, vol 3, 448, (1970). Our result disagrees with 
that of Silin (his eq.(3.12)) but argees with that obtained from the equations 
derived by Leggett. 

\end{thebibliography}
 
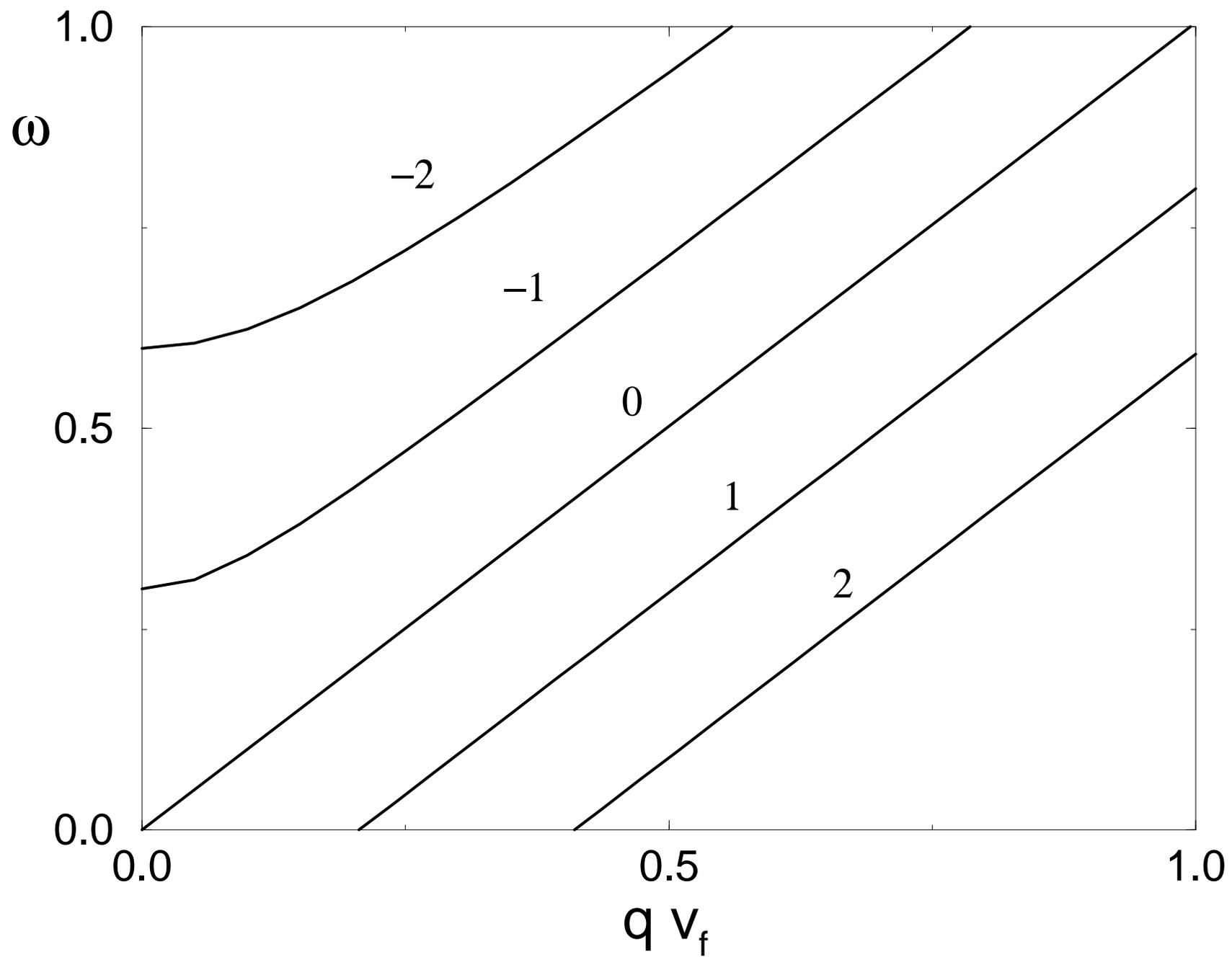
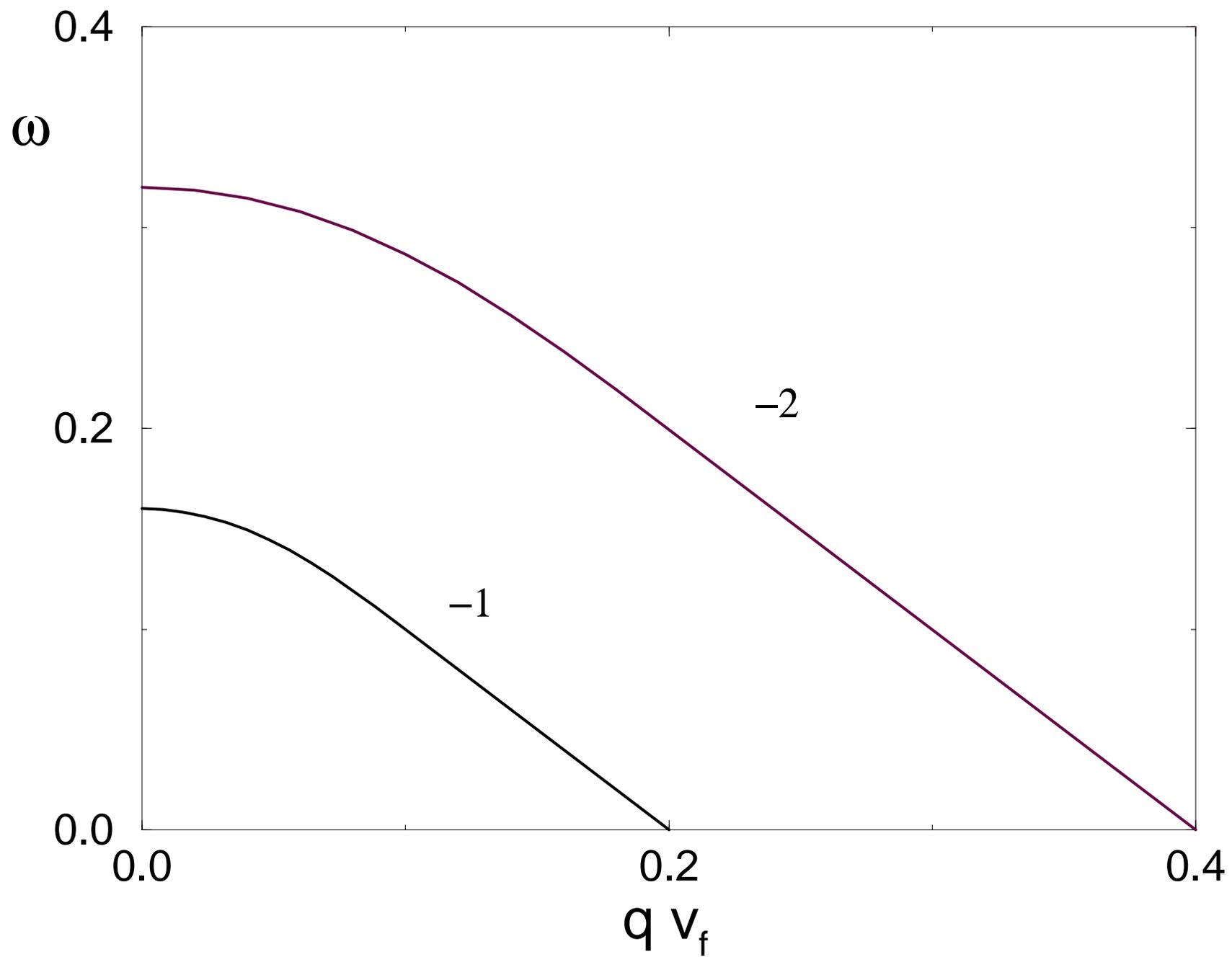
\end{document}